\documentclass[sigconf]{acmart}
\usepackage{subcaption}
\usepackage{graphicx}
\usepackage{float}
\usepackage{xcolor}
\usepackage{xspace}
\usepackage{environ}
\usepackage{framed}
\usepackage{multirow} 
\usepackage{algorithm}
\usepackage{fancyvrb}      
\usepackage{listings}
\usepackage[shortlabels]{enumitem}

\setlist[enumerate]{leftmargin=0.7cm,topsep=0.5mm}
\setlist[itemize]{leftmargin=0.5cm,topsep=0.5mm}

\usepackage[capitalise]{cleveref}
\newcommand\system{Kishuboard\xspace}
\newcommand\nrestart{Na\"iveRestart\xspace}
\newcommand\ncontinue{Na\"iveContinue\xspace}

\definecolor{RedColor}{HTML}{e66101}
\definecolor{OrangeColor}{HTML}{fdb863}
\definecolor{YellowColor}{HTML}{fec44f}
\definecolor{BlueColor}{HTML}{b2abd2}
\definecolor{PurpleColor}{HTML}{5e3c99}
\definecolor{Ours0Color}{HTML}{ABDDA4}
\definecolor{Ours16Color}{HTML}{72C166}
\definecolor{Ours32Color}{HTML}{38A528}
\definecolor{PrestoColor}{HTML}{999999}
\definecolor{PostgresColor}{HTML}{6D6D6D}
\definecolor{GreenColor}{HTML}{38A528}
\definecolor{YellowColor}{HTML}{ffb570}
\definecolor{BlueColor}{HTML}{7081ff}
\definecolor{PinkColor}{HTML}{ffb0c2}
\definecolor{ComputeColor}{HTML}{ffb0c2}
\definecolor{ReadColor}{HTML}{cf3457}
\definecolor{WriteColor}{HTML}{ffb570}
\definecolor{vintagegreen}{HTML}{ABDDA4}
\definecolor{OursColor}{HTML}{38A528}
\definecolor{GreedyColor}{HTML}{7081ff}
\definecolor{RandomColor}{HTML}{ffb570}
\definecolor{NoneColor}{HTML}{6D6D6D}
\definecolor{Redborder}{HTML}{805861}
\definecolor{Greenborder}{HTML}{384180}
\definecolor{Blueborder}{HTML}{566F52}
\definecolor{Greyborder}{HTML}{4D4D4D}
\definecolor{Lightgrey}{HTML}{dadada}
\definecolor{ExampleColor1}{HTML}{7081ff}
\definecolor{ExampleColor2}{HTML}{ffb0c2}
\definecolor{Lightred}{HTML}{ffb09c}
\definecolor{Lightblue}{HTML}{b8e2f2}
\definecolor{FlagColor}{HTML}{CCCCCC}
\definecolor{vintageblue}{HTML}{7081ff}
\definecolor{vintagered}{HTML}{ffb0c2}
\definecolor{NoOptColor}{HTML}{264653}
\definecolor{LRUColor}{HTML}{777777}
\definecolor{RandomColor}{HTML}{2a9d8f}
\definecolor{GreedyColor}{HTML}{e9c46a}
\definecolor{HeuristicColor}{HTML}{f4a261}
\definecolor{SCColor}{HTML}{e76f51}
\definecolor{AllColor}{HTML}{CCCCCC}
\definecolor{SAColor}{HTML}{ffb0c2}
\definecolor{SeparatorColor}{HTML}{9b5de5}
\definecolor{BlueColor}{HTML}{0081a7}
\definecolor{cAmain}{HTML}{e66101}
\colorlet{cAlight}{cAmain!25}
\colorlet{cAlightlight}{cAmain!5}
\definecolor{cBmain}{HTML}{5e3c99}
\colorlet{cBlight}{cBmain!25}
\definecolor{cCmain}{HTML}{b2abd2}
\colorlet{cClight}{cCmain!25}
\definecolor{cDmain}{HTML}{fdb863}
\colorlet{cDlight}{cDmain!25}
\definecolor{cZmain}{HTML}{030303}  
\colorlet{cZlight}{cZmain!25}
\colorlet{cZlightlight}{cZmain!5}
\colorlet{cZlightlightlight}{cZmain!1}
\definecolor{cPositivemain}{HTML}{2ca02c}  
\colorlet{cPositivelight}{cPositivemain!25}
\definecolor{cNegativemain}{HTML}{d62728}  
\colorlet{cNegativelight}{cNegativemain!25}

\usepackage{multirow}  
\usepackage{algorithm}
\usepackage{xcolor}      
\usepackage{fancyvrb}

\newcommand{\fix}[1]{{#1}}

\makeatletter
\apptocmd\@maketitle{{\myfigure{}\par}}{}{}
\makeatother

\copyrightyear{2025}
\acmYear{2025}
\acmConference[CHI '25]{CHI Conference on Human Factors in Computing Systems}{April 26-May 1, 2025}{Yokohama, Japan}
\acmBooktitle{CHI Conference on Human Factors in Computing Systems (CHI '25), April 26-May 1, 2025, Yokohama, Japan}\acmDOI{10.1145/3706598.3714141}
\acmISBN{979-8-4007-1394-1/25/04}
\setcopyright{cc}
\setcctype{by}

\copyrightyear{2025}
\acmYear{2025}
\setcopyright{rightsretained}
\acmConference[CHI EA '25]{Extended Abstracts of the CHI Conference on Human Factors in Computing Systems}{April 26-May 1, 2025}{Yokohama, Japan}
\acmBooktitle{Extended Abstracts of the CHI Conference on Human Factors in Computing Systems (CHI EA '25), April 26-May 1, 2025, Yokohama, Japan}\acmDOI{10.1145/3706599.3719928}
\acmISBN{979-8-4007-1395-8/2025/04}

\begin{document}

\title[Large-scale Evaluation of Notebook Checkpointing with AI Agents]{Large-scale Evaluation of Notebook Checkpointing with\\ AI Agents}






\author{Hanxi Fang}
\affiliation{%
  \institution{University of Illinois Urbana-Champaign}
  \city{Urbana}
  \state{Illinois}
  \country{USA}
}
\email{hanxif2@illinois.edu}

\author{Supawit Chockchowwat}
\affiliation{%
  \institution{University of Illinois Urbana-Champaign}
  \city{Urbana}
  \state{Illinois}
  \country{USA}
}
\email{supawit2@illinois.edu}

\author{Hari Sundaram}
\affiliation{%
  \institution{University of Illinois Urbana-Champaign}
  \city{Urbana}
  \state{Illinois}
  \country{USA}
}
\email{hs1@illinois.edu}

\author{Yongjoo Park}
\affiliation{%
  \institution{University of Illinois Urbana-Champaign}
  \city{Urbana}
  \state{Illinois}
  \country{USA}
}
\email{yongjoo@illinois.edu}

\begin{abstract}
Saving, or checkpointing, intermediate results during interactive data exploration can potentially boost user productivity. 
However, existing studies on this topic are limited, as they primarily rely on small-scale experiments with human participants---a fundamental constraint of human subject studies. 
To address this limitation, we employ AI agents to simulate a large number of complex data exploration scenarios,
including revisiting past states and branching into new exploration paths. 
This strategy enables us to accurately assess the impact of checkpointing 
	while closely mimicking the behavior of real-world data practitioners. 
Our evaluation results, involving more than 1,000 exploration paths and 2,848 executed code blocks, 
    show that a checkpointing framework for computational notebooks 
        can indeed enhance productivity
    by minimizing unnecessary code re-executions
        and redundant variables/code.
\end{abstract}

\maketitle

\section{Introduction}


Checkpointing computational notebooks can improve
    user productivity~\cite{hanxichi}.
Specifically, notebook systems like Jupyter~\cite{jupyter}, Colab~\cite{colab}, R Markdown~\cite{rmarkdown} let data scientists
    run a code block one at a time for
        analyzing tabular data, training machine learning models,
            visualizing results, etc.
By checkpointing intermediate code/data,
    users can undo undesirable executions,
        explore alternative hypotheses,
            and restore from crashes.
A recent work~\cite{hanxichi} proposes a checkpointing interface for computational notebooks
    and demonstrates its productivity benefits.

Unfortunately,
    the existing evaluation is limited to employing a handful of human subjects
        working on a fixed set of tasks.
While those participants are allowed
    to freely explore,
    the study remains constrained by the number of participants, 
        the variety of tasks, and the duration of the study.
We recognize that these limitations are inherent to 
    human subject studies: the limitations are difficult to overcome without adopting an entirely different approach.

In this work,
    we tackle them by devising an AI agent-based strategy.
Our observation is that AI agents can generate high-quality code akin to that produced by real data scientists, and exhibit iterative refinement behaviors that closely resemble the testing, debugging, and code improvement processes typically employed by human practitioners.
Specifically,
    we employ a pre-trained AI agent (i.e., ChatGPT 4o)
    to simulate real-world data scientists who explore data
        \textbf{with and without} a notebook checkpointing tool.
For each of the scenarios,
    the only independent variable is the assistance offered by the checkpointing tool;
    however, the actual variables residing in sessions
        and also elapsed times
        vary significantly.

This ``late-breaking work'', as a sequel to the recent research paper~\cite{hanxichi},
    expands the current state-of-the-art in two significant aspects.
First, we systematically evaluate the effectiveness of the new code+data space checkpointing framework
    with hundreds of data exploration scenarios. 
    This work is the first that evaluates data science checkpointing frameworks~\cite{li2023elasticnotebook,kishu,chex,wang2022diff}
        in such a large scale.
Second, we reason why our agent-based approach is valid,
    ensuring two types of consistencies:
        consistency of generated code and consistency of branching strategy.

\section{Background}
\label{sec:background}

We overview notebook checkpointing frameworks (\cref{sec:background:checkpointing})
    and discuss the limitations of the existing evaluation (\cref{sec:existing_eval}).

\begin{figure*}[t]
\includegraphics[width=0.9\linewidth]{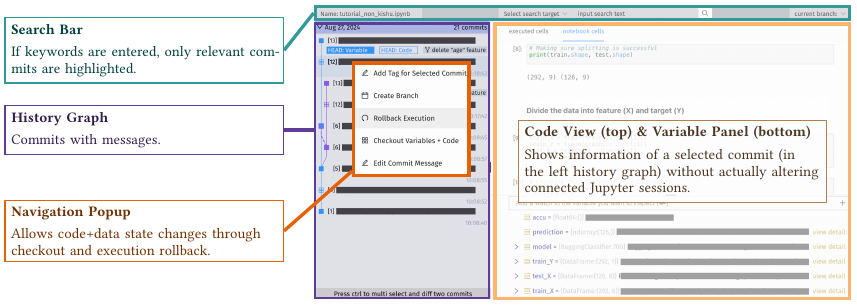}
\vspace{-2mm}
\caption{
Background about \system.
The history graph (purple box) shows past commits.
The code and variable panes (yellow box) display
    the information of a selected commit.
From any past commit,
    users can load data only (i.e., execution rollback)
        or load both code and data (i.e., checkout)
    using the navigation popup (red box).
The image is reproduced with the original authors' permission.
}
\Description{
 This figure shows Kishuboard User Interface

The interface includes:

Search Bar: Filters commits based on keywords.
History Graph: Displays past commits with code and data states.
Navigation Popup: Allows switching between code+data states (rollback or checkout).
Code View & Variable Panel: Shows details of the selected commit without changing the active Jupyter session.
This interface helps users manage and explore different versions of computational notebooks effectively.
}
\label{fig:ui:kishuboard}
\end{figure*}


\subsection{Notebook Checkpointing}
\label{sec:background:checkpointing}

Computational notebooks (e.g., Jupyter, Colab) are designed around a linear, cell-by-cell execution model, even though real-world data science exploration is often iterative and non-linear. Practitioners commonly revisit older states, branch the code to try different alternatives, and switch among branches.~\cite{rule2018exploration,kery2019towards} To accommodate branching, they either start a new notebook for each path~\cite{chattopadhyay2020s} or mix branch-specific code cells in a single notebook~\cite{liu2019understanding,subramanian2019supporting,kery2018story}. Both approaches are time-consuming (re-running shared cells) and cognitively taxing (tracking which cells belong to each branch).

Inspired by code version control systems~\cite{gitbook}, \system~\cite{hanxichi}
addresses these issues by introducing \textbf{two-dimensional code+data checkpointing} for computational notebooks (see \cref{fig:ui:kishuboard} for its user interface). Unlike one-dimensional version control tools that track only code revisions, \system also version-controls the evolving “variable state in the kernel” or “data state” produced by each cell’s execution. Based on that, users can \textbf{checkout} the code and data state of notebook to any previous point and start exploring a new branch from that point. This feature, together with \system's UI that visualizes the branched structure of exploration similar to Git GUI~\cite{git_gui_clients}, enables fast, nonlinear data science exploration and reduces user's cognitive burden of managing multiple exploration branches simultaneously.

\subsection{Existing Evaluation of Notebook Checkpointing and its Limitations}
\label{sec:existing_eval}

The existing evaluation is based on human subject studies~\cite{hanxichi}. Specifically, to evaluate whether notebook checkpointing improves data science productivity, a user study was conducted with 20 student participants randomly split into experimental (with \system to checkout states) and control (without \system) groups. Both groups performed the same notebook-based tasks---building models, branching workflows, reporting metrics, retrieving previous variable states, debugging, and recovering from system crashes. While data practitioners may write their own code, participants in this study chose from a set of pre-written code snippets (with the option to modify them), ensuring that notebook checkpointing was the primary independent variable. The study measured each participant’s task completion time and gathered feedback through surveys, then compared performance between the two groups.

Evaluation using traditional user study has several limitations:
\begin{enumerate}
  \item As it relies on a small group of participants, its statistical power is limited.
    \item The short study duration restricts participants’ ability to engage in extended exploration. Even if they are allowed to branch and iterate, the limited time reduces opportunities to test multiple hypotheses or recover from mistakes.
  \item The task steps and reference code snippets are provided, 
    limiting diverse exploration.
\end{enumerate}

These limitations are inherent to  resource-constrained human-subject user study. 
To address them under the user-study paradigm, one would need to recruit more participants, extend the study duration, and ensure uniform expertise in data science and Python coding---all of which may involve substantial costs.

\begin{figure*}[t]
\includegraphics[width=0.9\linewidth]{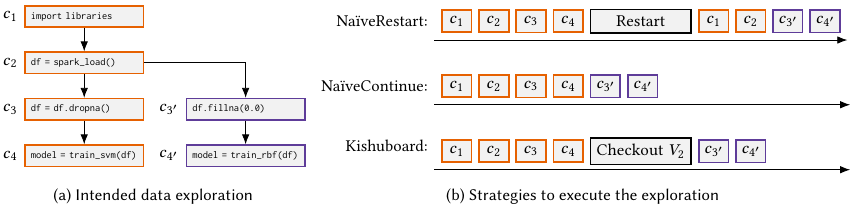}
    \caption{\fix{A toy example of user's intended data exploration and strategies to execute it. \nrestart repetitively executes cells $c_1$ and $c_2$. \ncontinue executes new cells ($c_{3'}$ and $c_{4'}$) without any kernel restart. \ncontinue may lead to branch interferences, for example, N/A values were already dropped by $c_3$, making data imputation in $c_{3'}$ ineffective. 
    \system restores checkpointed data to explore a new path,
        thus removing the repetitive work and preventing potential branch interferences.}}
    \label{fig:llm_group_example}
\end{figure*}

\section{AI-Agent-Based Experiment Design}
\label{sec:design}
We tackle the limitations mentioned in \cref{sec:existing_eval} 
    using AI agents.
    As AI agents can generate diverse code
        in a short amount of time, compared to humans, 
    it allows us to evaluate \system's effectiveness 
        at a much larger scale
        in terms of code diversity at a relatively small cost.
    We first describe our experiment design in this section.
    Then, we examine its scientific validity in \cref{sec:validation}.
    
    \paragraph{Simulating Branched Data Exploration} We use a large language model (LLM),
        i.e., ChatGPT-4o~\cite{achiam2023gpt}, 
    to iteratively generate code cells based on a generic task specification. We first provide the task: ``My data is in \texttt{file\_name}. The columns and their meanings are \texttt{column\_names\_and\_descriptions}. I want to build a model to predict \texttt{requirements}.'' then prompt the LLM to outline a step-by-step plan. Afterward, we prompt the agent to generate code cells and execute the generated code according to the plan step by step. Specifically, the execution results and code for previous steps are fed to the agent when generating the code for next step. 

    Data scientists often revisit previous explored phases~\cite{liu2019understanding,kery2018story,guo2013workflow,raghunandan2023code,koop2017dataflow,hohman2020understanding}. To simulate such a behavior, we randomly select a past cell to revisit and ask the LLM to generate alternative code cells to continue from the selected cell. Each time creates a new branch of exploration. Whenever a runtime error occurs, the LLM can retry twice to self-correct according to the error message. If the LLM fails to fix the error, we checkout randomly and start a new branch.
    \cref{sec:appendix:prompts} presents concrete examples
        of our prompts and generated code by the LLM.
    
    We repeat the generation process 10 times to simulate \textbf{10 exploration sessions} of data scientists exploring data independently where each session explores \textbf{100 branches}. The dataset is the 32~KB Titanic dataset downloaded from Kaggle~\cite{titanicdata}. Overall, the LLM generated \textbf{2,848} cells in this study. 

    To test \system's performance with memory-intensive tasks, we also use the Spotify podcast dataset (i.e., 45 MB and 450 MB)~\cite{spotifydata} to simulate 3 exploration sessions with each exploring 3 branches.
    
    \paragraph{Baselines and \system}
    Given an exploration session, we test the following existing baseline strategies and \system to support branched data exploration (see \Cref{fig:llm_group_example} for an example).
    (1) \textbf{\nrestart}: To explore a new branch, \nrestart restarts the kernel and re-executes common cells to restore the selected past state. This strategy follows some user's behaviour of starting a new notebook for a new exploration path~\cite{chattopadhyay2020s}.
    (2) \textbf{\ncontinue}: Instead, \ncontinue appends new cells to the current notebook and continues execution from the current kernel state
        without any restart. 
    It mimics a common practice to mix code from different cells within the same notebook~\cite{liu2019understanding,subramanian2019supporting,kery2018story}. \ncontinue can introduce both explicit errors (e.g., operating on columns dropped in an earlier branch, thus observing execution errors) or implicit errors (e.g., unknowingly working on a dataframe updated by an earlier branch).
    (3) \textbf{\system}: This approach leverages \system to check out the code+data version to the latest common history state and continue execution from there.

    \paragraph{Metrics}

    In addition to performance metrics (i.e., execution times), this experiment is also interested in measuring correctness and exploration complexity using the following metrics.
    (1) \textbf{End-to-end time to explore all branches}: This metric measures the time elapsed to execute and checkpoint/checkout code for all the generated branches.
    It focuses solely on system efficiency during execution and state management (if any) as the time for AI-agents to generate code is neglected.
    (2) \textbf{Number of branch interferences}:
    In the real world, a \emph{branch interference} occurs when a user reuses a modified variable assuming it was unmodified. Unfortunately, it may lead to misleading exploration outcomes which are hard to detect and debug.
    To study this type of mistake, we count the number of times a variable is incorrectly accessed across branches.
    For example, in \cref{fig:llm_group_example}, cell $c_3$ from the first branch drops rows with missing values in \texttt{df}, but with naively continue executing, cell $c_{3'}$ would ineffectively impute missing-but-already-dropped values which is a branch interference.
    We detect such branch interference by analyzing variable access and modification in each cell as well as manually verifying each detected branch interference.
    (3) \textbf{Peak number of variables}: This metric tracks the maximum number of variables in the kernel's namespace. A higher number indicates a greater memory burden and increased cognitive overhead for the user to remember and manage the variables.
    (4) \textbf{Peak number of cells}: This metric tracks the maximum number of cells in the notebook at any given time. A larger number of cells indicates a messier notebook cell organization, leading to difficulties working with the notebook and cognitive burden.

\begin{figure*}[ht]
\centering
\includegraphics[width=0.9\linewidth]{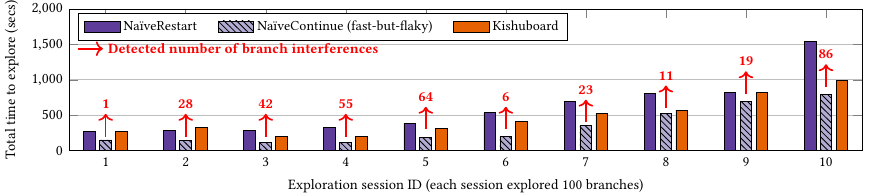}
\caption{\fix{End-to-end execution time for \system and baseline methods. We generated \textbf{1000} branches of code using LLM-Agent, divided into 10 exploration sessions with 100 branches each. The sessions are sorted in ascending order by \nrestart time. \ncontinue method is the fastest in terms of execution time, as it only runs newly added cells without checkpoint or checkout overhead. However, it is faulty, often producing incorrect results that do not trigger explicit errors, which may require significant debugging time. The additional time for the \system group is due entirely to checkpointing and checkout overhead, with the worst-case average overhead being just 2 seconds per branch. The red annotations indicate the number of implicit incorrect results for each session.}}

\vspace{-2mm}
\label{fig:e2e_time}
\end{figure*}

\section{Validation of the Experiment Design}
\label{sec:validation}

In this section, we aim to justify the validity of AI-agent-based experiment design. Specifically, we discuss the similarity between AI agents and human behavior in terms of code generation. We also explain why the experiment setting effectively measures the efficiency gain of notebook checkpointing from a causal perspective.

\paragraph{Similarity to Human Exploration Behavior}
Our decision to employ an AI agent for code generation and exploration is founded on recent research that demonstrates the capability of large language models (LLMs) to produce code (including data science code) and iterative refinement patterns that closely resemble those of human programmers~\cite{wei2022chain,hong2024data,guo2024ds, xu2022systematic}. Concretely, LLMs has been shown to:
\begin{enumerate}
\item \textbf{Generate high-quality data science code:} LLMs can generate high-quality code for data science tasks. For example, recent research shows LLM-based agent can achieve accuracy of 94.9\% on certain data analysis task benchmarks~\cite{hong2024data} and get mean ROC AUC performance of 0.82 on some feature engineering benchmarks~\cite{hollmann2024large}.

\item \textbf{Iteratively refine code:} Research in LLM agents~\cite{wei2022chain,yao2022react,shinn2024reflexion} shows that LLMs can parse feedback from external contexts to improve outputs. In our setup, the AI agent generates code step-by-step, receives execution feedback at each stage. Like a human data scientist, the AI agent can iteratively test, debug, and refine their work.
\end{enumerate}
While the AI agent is not subject to the same cognitive constraints as a human (e.g., short-term memory limitations or domain biases), these overlaps in behavior patterns provide reasonable assurance that our experimental tasks and the corresponding branching strategies align with real-world exploration scenarios. Moreover, as many data scientists use AI for coding assistance~\cite{nijkamp2022codegen, github_copilot,fried2022incoder}, their code may employ part of AI-agent-generated code.

\paragraph{Causal Perspective on Measuring Efficiency Gains}
A key rationale for our experiment is to measure, in a controlled manner, how much efficiency researchers can gain by using a notebook checkpointing mechanism (i.e.,\system). From a causal inference perspective, the central challenge is to isolate the effect of the checkpointing strategy from other factors, such as task difficulty or participants' skill level. Our design addresses this issue in two ways:

\begin{enumerate}
    \item \textbf{Consistency of Generated Code:} By using the same generated code across different checkpointing strategies, we ensure that our experiment holds task complexity and baseline code constant. Because the generated code and data do not change, any observed differences in runtime behavior or outcomes must stem from the checkpointing mechanism itself.
    
    \item \textbf{Consistency of Branching Strategy:} We introduce random selection to revisit previous cells, mirroring real-world data exploration where users backtrack to earlier steps~\cite{kery2018story,liu2019understanding}. Each branch emerges from the same generative process but differs in terms of whether it restarts the kernel, continues with the existing kernel state, or checks out a prior code-data version. This design ensures that all strategies face the same exploration tasks, allowing a fair comparison of performance.
\end{enumerate}




   \section{Experiment Result and Discussion}
    \label{sec:llm-result}

    Using the AI-agent-based methodology described earlier, we present and discuss the experiment results in this section in terms of correctness, efficiency, and notebook simplicity. In summary, we get the following experiment results:
     \begin{enumerate}
        \item \system speeds up data exploration: up to $36\%$ faster in both compute-intensive/light tasks, than naively restarting and re-execution. On average, data exploration with \system is $23\%$ faster in compute-intensive tasks and 
        $15\%$ faster in compute-light tasks
            in terms of execution times. 
        \item \system prevents incorrect results 
            that often occur when users na\"ively explore alternative paths
                by appending cells and/or reusing variables.
        \item \system maintains a clean notebook with only cells and variables in the current branch, which is 
            significantly more effective than
            keeping track of branched structures with (almost) redundant code and variables.
        \item \system's space overhead---checkpoint sizes---is
            small.
    \end{enumerate}
    
    \paragraph{\system Speeds up Data Exploration} 
    We compare end-to-end times to explore using different strategies. \Cref{fig:e2e_time} shows that \system enables faster data explorations than \nrestart does. In Session 10, \system would save the user about 10 minutes of real-time or 36\% faster than \nrestart. As the only exception, \system is slightly slower than \nrestart in Session 2 because cells before the branching point are all lightweight, making recomputation faster than using recovering from checkpoints.
    \ncontinue is expectedly faster than \system as \ncontinue overwrites the existing kernel state by running only the additional cells (for a new exploration path);
    however, \ncontinue causes significant issues 
        such as branch interferences and incorrect results as described below.


\begin{figure*}[t]
\includegraphics[width=0.9\linewidth]{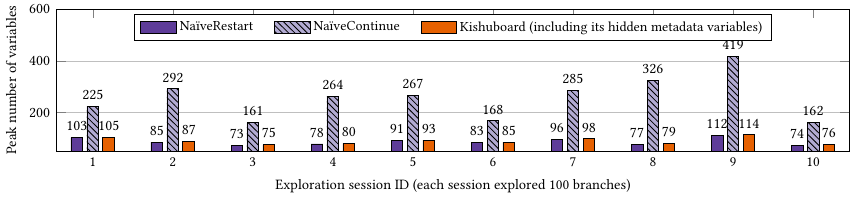}
\caption{\fix{The peak number of kernel variables for each session during exploration. 
Smaller numbers of variables may be preferred for easier understanding.
\ncontinue produces excessive variables, increasing cognitive load to keep track of variables across branches. \system has exactly two more variables than \nrestart for user-invisible metadata.}}

\vspace{-2mm}
\label{fig:complexity_var}
\end{figure*}

    
    \paragraph{\system Ensures Correctness} Furthermore, \system guarantees correctness (\Cref{fig:e2e_time}) by recovering exact data states from checkpoints.
    In contrast, while the alternative method \ncontinue is the fastest, it is error-prone. In particular, we detect 355 branch interferences which are hard to detect and debug manually.
    For example, in Session 9, Branch 5 modifies a variable \texttt{X\_test} by resampling from the original dataset and re-assigning it to the same variable name. Later on, Branch 7 evaluates its model based on \texttt{X\_test} which leads to a misleading accuracy because the resampled \texttt{X\_test} overlaps with the training data. 
    Unsurprisingly, \nrestart also guarantees correctness but incurs computation costs to re-execute cells common across branches.

    \paragraph{\system Maintains Clean Code and Data States} To assess notebook complexity, we measure the peak number of variables during exploration, as shown in \Cref{fig:complexity_var}. Compared to \nrestart and \system, \ncontinue results in significantly more variables, 
    because \ncontinue mixes data states from multiple exploration branches. We also measure the peak number of cells in the notebook during exploration and find that the \ncontinue sessions have at least $24\times$ more cells than the other two groups, as it appends code from different branches to the same notebook.
    Consequently, the user of \ncontinue would experience an increased cognitive load to manage code cells across branches and remember variables.

    
    \paragraph{Simulating Memory-intensive Tasks} Using the AI agent to explore both 45~MB and 450~MB Spotify dataset~\cite{spotifydata},
    our results presented in \cref{table:large_dataset} show that
    \system can checkpoint variables efficiently without requiring extensive IO overhead or storage resources. For the 45~MB dataset, the total checkpoint size was 56~MB, and for the 450~MB dataset, it was 358~MB---smaller than the dataset itself in a CSV format.
    Furthermore, \system accelerates data exploration in both memory-light and memory-intensive scenarios, demonstrating the scalability of the method.

    We observed that during experiments with larger datasets, \system occasionally fails with ``\texttt{OverflowError: BLOB longer than INT\_MAX bytes}'' due to a limitation in the existing backend implementation~\cite{kishu}.
    Because this issue is an implementation flaw and not inherent to the technique, we believe that it could be addressed by chunking data before writing to storage.

\begin{table*}[t]
\caption{\fix{End-to-end execution time for exploring three randomly generated branches on small (45 MB) and large (450 MB) datasets. ``Time'' denotes the duration from executing the first cell of the first branch to completing the last cell of the final branch. ``\# of Cells'' and ``\# of Variables'' indicate the \textbf{peak} number of cells in the notebook and the peak number of variables in the kernel, respectively. 
\system outperforms \nrestart in efficiency for both dataset sizes, demonstrating the scalability of the method. Additionally, \system requires only 56 MB of storage for checkpoints on the small dataset and 358 MB on the large dataset throughout the exploration.
While \ncontinue is faster than \system, 
    as observed in the previous experiments,
    \ncontinue often causes hard-to-detect errors (\cref{fig:e2e_time})
        and creates significantly more variables(\cref{fig:complexity_var}) and cells.
}}
\label{table:large_dataset}

\centering
\small
\begin{tabular}{lrrrrrr}
\toprule
\multirow{2}{*}{\textbf{Method}}
    & \multicolumn{3}{c}{Small Dataset (45 MB)} & \multicolumn{3}{c}{Large Dataset (450 MB)}\\
    & \textbf{Time (secs)} & \textbf{\# of Cells} & \textbf{\# of Variables}  & \textbf{Time (secs)} & \textbf{\# of Cells} & \textbf{\# of Variables} \\
\midrule
\nrestart & 38 & 9 & 63 & 271 & 9 & 63 \\
\ncontinue & 18 & 14 & 69 & 141 & 14 & 69 \\
\system & 24 & 9 & 65 & 169 & 9 & 65 \\
\bottomrule
\end{tabular}
\end{table*}

\section{Limitations of Experiment Setup and Result}
While our AI-agent-based evaluation provides a large-scale and systematic analysis of notebook checkpointing, several limitations must be acknowledged. These primarily stem from the differences in behavior between AI agents and human practitioners.

\paragraph{AI Agents Do Not Replicate How Humans Iteratively Explore Branches} The simulation of 100 randomly generated branches within a 10-cell notebook may not fully capture real-world data science workflows. While AI agents randomly revert to past commits, human users make such decisions based on logical reasoning or execution results. Furthermore, human users vary in exploration depth, leading to different numbers of executed cells per branch. This discrepancy may influence our findings, particularly in assessing the efficiency of different checkpointing strategies. For example, 
if humans always choose to checkout to a very
early stage, then the efficiency gap between \nrestart and \system may be smaller.

\paragraph{Unlike AI Agents, Humans Can Mitigate Variable Overwrites} The drawbacks observed in the \ncontinue group may not be as severe in real-world scenarios. Experienced users often mitigate variable overwrites by explicitly renaming or copying variables when switching between branches. As a result, the frequency of branch interferences leading to execution errors may be lower than what our AI-agent simulations suggest. Moreover, while AI agents append all executed cells into a single notebook, real users often refactor their workflow by removing obsolete cells or reorganizing their notebooks dynamically. However, it is important to note that these manual interventions take time—time that \system aims to save by automating refactoring and state management.

\paragraph{AI Agents Do Not Experience Cognitive Load} In real-world scenarios, users must actively navigate commit histories and decide which version to restore using \system. Similarly, those in the \ncontinue and \nrestart groups would spend additional time figuring out how to organize cells or new notebooks. These interactions, which require human judgment and effort, are not fully accounted for in our AI-driven evaluation.

Given these factors, while our AI-agent-based approach offers valuable insights into checkpointing strategies at scale, complementary user studies are needed to validate its real-world applicability.

\section{Conclusion}
In this study, we demonstrate the efficacy of using AI agents to evaluate notebook checkpointing systems at scale, addressing limitations inherent in traditional human-subject studies. Traditional user studies are constrained by limited participant pools, short durations, and the substantial costs of ensuring uniform skill validation, which collectively hinder the scale and diversity of such evaluations. In contrast, AI agents offer a scalable alternative by generating diverse code in a short amount of time, mimicking real-world exploratory behaviors, and producing high-quality code. By simulating diverse and complex data exploration scenarios—encompassing a total of 1,000 branches— we assessed the performance of Kishuboard, a novel code+data checkpointing framework, against baselines. The results show that Kishuboard significantly enhances exploration efficiency, reduces cognitive burden, and prevents branching errors. Our work is the first of its kind to demonstrate the utility of AI agents in evaluating notebook checkpointing systems.

\bibliographystyle{ACM-Reference-Format}
\bibliography{refs}
\clearpage

\appendix
\section{AI-Agent Prompts and Generated Notebooks}
\label{sec:appendix:prompts}

\DefineVerbatimEnvironment{SVerbatim}{Verbatim}{frame=single}
\lstset{
  breaklines=true,
  basicstyle=\ttfamily\footnotesize, 
  frame=single,               
  showstringspaces=false,
  columns=fullflexible,
  breakautoindent=false,breakindent=0ex
}
\subsection{Prompts}

We first prompt ChatGPT-4 with the following prompts to generate steps it is planing to take, here's a real example:

\begin{lstlisting}
"My dataset is in top_podcasts.csv, it has header. It has the following fields:<all fields name in the dataset>. I want to build a model to track how podcast rankings change over time to identify emerging trends. 

It will be an interactive data science process using notebook.Tell me the steps to do this (without code). The steps should include some computation intensive ones such as hyperparameter tuning, model selection, neural network training, etc.

The step should be listed as:
There are XXX steps in total.
1. XXX
2. XXX
\end{lstlisting}

Next, we'll ask ChatGPT-4 to generate code step by step, along with the result of the last step feeded, here's an example:

\begin{lstlisting}
"The output for this code cell is: <execution output>. Please generate the code for next step <step number>.
\end{lstlisting}

When the agent's output code contains error when executing, we give it 2 chances to self-correct, with the following prompt:

\begin{lstlisting}
The output of the given code is: {output}, please help me debug it by generating the correct cell code.
\end{lstlisting}
Note that, when interacting with ChatGPT, we also include the history messages in the current prompt. 
When checking out, we also checkout the messages to the right point, so that ChatGPT only has the information about previous steps before the current step.

\subsection{Example Generated Code}
Here's a real example of the generated code branch, with step generated one by one as separate cells:

\begin{lstlisting}
# Step 1: Data Loading

import pandas as pd

# Load the dataset
file_path = 'titanic.csv'
titanic_data = pd.read_csv(file_path)

# Display the first few rows of the dataset
titanic_data.head()
\end{lstlisting}
\begin{lstlisting}
# Step 2: Data Exploration

import matplotlib.pyplot as plt
import seaborn as sns

# Display basic information about the dataset
titanic_data.info()

# Check for missing values
missing_values = titanic_data.isnull().sum()
print("Missing values in each column:\n", missing_values)

# Visualize the distribution of the 'Survived' variable
sns.countplot(x='Survived', data=titanic_data)
plt.title('Survival Counts')
plt.xlabel('Survived (0 = No, 1 = Yes)')
plt.ylabel('Count')
plt.show()

# Visualize the distribution of 'Age' with respect to 'Survived'
plt.figure(figsize=(12, 6))
sns.histplot(data=titanic_data, x='Age', hue='Survived', multiple='stack', bins=30)
plt.title('Age Distribution by Survival')
plt.xlabel('Age')
plt.ylabel('Count')
plt.show()

# Visualize the relationship between 'Pclass' and 'Survived'
sns.countplot(x='Pclass', hue='Survived', data=titanic_data)
plt.title('Survival Counts by Passenger Class')
plt.xlabel('Passenger Class')
plt.ylabel('Count')
plt.show()

# Pairplot to visualize relationships between features
sns.pairplot(titanic_data, hue='Survived', diag_kind='kde')
plt.show()
\end{lstlisting}
\begin{lstlisting}
# Step 3: Data Preprocessing

# Make a copy of the original dataset
titanic_data = titanic_data.copy()

# Handle missing values for 'Age' by filling with the median age
titanic_data['Age'].fillna(titanic_data['Age'].median(), inplace=True)

# Check again for missing values to confirm they have been addressed
missing_values_after = titanic_data.isnull().sum()
print("Missing values after preprocessing:\n", missing_values_after)

# Encode categorical variables using one-hot encoding for 'Embarked' and label encoding for 
'Sex'
titanic_data = pd.get_dummies(titanic_data, columns=['Embarked'], drop_first=True) 
titanic_data['Sex'] = titanic_data['Sex'].map({'male': 0, 'female': 1})  # Label encoding

# Drop columns that are not needed for model training
titanic_data.drop(['PassengerId', 'Name', 'Ticket'], axis=1, inplace=True)

# Display the preprocessed dataset
titanic_data.head()
\end{lstlisting}
\begin{lstlisting}
# Step 4: Feature Selection

# Import necessary libraries
from sklearn.model_selection import train_test_split
from sklearn.ensemble import RandomForestClassifier
from sklearn.feature_selection import SelectFromModel

# Define the features and target variable
X = titanic_data.drop('Survived', axis=1)
y = titanic_data['Survived']

# Split the dataset into training and testing sets
X_train, X_test, y_train, y_test = train_test_split(X, y, test_size=0.2, random_state=42)

# Train a Random Forest classifier to evaluate feature importance
rf_model = RandomForestClassifier(random_state=42)
rf_model.fit(X_train, y_train)

# Get feature importances
importances = rf_model.feature_importances_

# Create a DataFrame for feature importances
feature_importance_df = pd.DataFrame({'Feature': X.columns, 'Importance': importances})
feature_importance_df = feature_importance_df.sort_values(by='Importance', ascending=False)

# Display feature importances
print("Feature Importances:\n", feature_importance_df)

# Select features that are important (e.g., importance >= 0.05)
selected_features = feature_importance_df[feature_importance_df['Importance'] >= 0.05]
['Feature'].tolist()
print("Selected Features:\n", selected_features)

# Create a new dataset with selected features
X_train_selected = X_train[selected_features]
X_test_selected = X_test[selected_features]
\end{lstlisting}
\begin{lstlisting}
# Step 5: Data Splitting

# As we have already split the data in the previous step,
# we'll confirm the split and prepare for modeling.

# Display the shape of the training and testing sets
print("Training set shape (features):", X_train_selected.shape)
print("Testing set shape (features):", X_test_selected.shape)
print("Training set shape (target):", y_train.shape)
print("Testing set shape (target):", y_test.shape)

# Optionally, you may want to reset the indices of the training 
# and testing sets for consistency
X_train_selected.reset_index(drop=True, inplace=True)
X_test_selected.reset_index(drop=True, inplace=True)
y_train.reset_index(drop=True, inplace=True)
y_test.reset_index(drop=True, inplace=True)

# Display the first few rows of the training set
print("First few rows of the training set (features):\n", X_train_selected.head())
print("First few rows of the training set (target):\n", y_train.head())
\end{lstlisting}
\begin{lstlisting}
# Step 6: Model Selection

# Import necessary libraries
from sklearn.linear_model import LogisticRegression
from sklearn.tree import DecisionTreeClassifier
from sklearn.ensemble import RandomForestClassifier
from sklearn.svm import SVC
from sklearn.neighbors import KNeighborsClassifier
from sklearn.metrics import accuracy_score, classification_report, confusion_matrix

# Define a list of models to evaluate
models = {
    'Logistic Regression': LogisticRegression(),
    'Decision Tree': DecisionTreeClassifier(random_state=42),
    'Random Forest': RandomForestClassifier(random_state=42),
    'Support Vector Machine': SVC(random_state=42),
    'K-Nearest Neighbors': KNeighborsClassifier()
}

# Initialize a dictionary to hold model performance
model_performance = {}

# Fit each model and evaluate its performance
for model_name, model in models.items():
    model.fit(X_train_selected, y_train)  # Fit the model
    y_pred = model.predict(X_test_selected)  # Make predictions
    
    # Calculate accuracy
    accuracy = accuracy_score(y_test, y_pred)
    model_performance[model_name] = accuracy
    
    # Print classification report and confusion matrix for each model
    print(f"Model: {model_name}")
    print(f"Accuracy: {accuracy:.4f}")
    print("Classification Report:\n", classification_report(y_test, y_pred))
    print("Confusion Matrix:\n", confusion_matrix(y_test, y_pred))
    print("-" * 50)

# Display performance summary for all models
print("Model Performance Summary:\n", model_performance)
\end{lstlisting}
\begin{lstlisting}
# Step 7: Hyperparameter Tuning

# Import necessary libraries for hyperparameter tuning
from sklearn.model_selection import GridSearchCV

# Define the model to tune (example: Random Forest)
model_to_tune = RandomForestClassifier(random_state=42)

# Define the hyperparameters to test
param_grid = {
    'n_estimators': [50, 100, 200],  # Number of trees in the forest
    'max_depth': [None, 10, 20, 30],  # Maximum depth of the tree
    'min_samples_split': [2, 5, 10],  # Minimum number of samples required to 
                                      #split an internal node
    'min_samples_leaf': [1, 2, 4]  # Minimum number of samples required to be at a leaf node
}

# Initialize GridSearchCV
grid_search = GridSearchCV(estimator=model_to_tune, param_grid=param_grid,
                           scoring='accuracy', cv=5, verbose=1, n_jobs=-1)

# Fit the model to the training data
grid_search.fit(X_train_selected, y_train)

# Get the best parameters and best score
best_params = grid_search.best_params_
best_score = grid_search.best_score_

print("Best Parameters from Grid Search:", best_params)
print("Best Cross-Validated Accuracy:", best_score)

# Optionally, we can also evaluate the best model on the test set
best_model = grid_search.best_estimator_
y_pred_best = best_model.predict(X_test_selected)

# Evaluate the best model on the test set
best_accuracy = accuracy_score(y_test, y_pred_best)
print(f"Best Model Test Accuracy: {best_accuracy:.4f}")
print("Classification Report of Best Model:\n", classification_report(y_test, y_pred_best))
print("Confusion Matrix of Best Model:\n", confusion_matrix(y_test, y_pred_best))
\end{lstlisting}
\begin{lstlisting}
# Step 8: Model Training

# As we have already identified the best model through hyperparameter tuning,
# we will now re-train this best model on the entire training dataset.

# Fit the best model on the entire training set
best_model.fit(X_train_selected, y_train)

# Make predictions on the test set
y_pred_final = best_model.predict(X_test_selected)

# Evaluate the final model's performance on the test set
final_accuracy = accuracy_score(y_test, y_pred_final)
print(f"Final Model Test Accuracy: {final_accuracy:.4f}")
print("Final Classification Report:\n", classification_report(y_test, y_pred_final))
print("Final Confusion Matrix:\n", confusion_matrix(y_test, y_pred_final))

# Optionally, save the model for future use
import joblib
joblib.dump(best_model, 'best_random_forest_model.pkl')
print("Best model saved as 'best_random_forest_model.pkl'.")
\end{lstlisting}

\end{document}